\documentclass[%
reprint,
superscriptaddress,
amsmath,amssymb,
aps,
pra,
floatfix,
]{revtex4-2}

\usepackage{bm}
\usepackage{mathtools}
\usepackage{graphicx}
\usepackage{mathrsfs}
\usepackage{color}
\usepackage{amsfonts}
\usepackage{times,txfonts}
\usepackage{nicefrac}
\usepackage{physics}
\usepackage{soul}

\usepackage[colorlinks=true,linkcolor=blue,urlcolor=blue,citecolor=blue,pdfusetitle]{hyperref}

\newcommand{\trp}{\mathsf{T}}
\newcommand{\ham}{H} % Hamiltonian
\newcommand{\tham}{\tilde\ham}
\newcommand{\ad}{a^\dag} % creation operator
\newcommand{\lx}{\lambda_x} % couplings
\newcommand{\ly}{\lambda_y}
\newcommand{\lc}{\lambda_c} % critical coupling
\newcommand{\iu}{\mathrm{i}} % mathematical constants
\newcommand{\eu}{\mathrm{e}}
\newcommand{\cov}{C} % covariance matrix
\newcommand{\sysj}{\mathsf{j}}
\newcommand{\sysx}{\mathsf{x}}
\newcommand{\sysy}{\mathsf{y}}

\newcommand{\sref}[2]{\hyperref[#1]{Fig.~\ref*{#1}#2}}
\newcommand{\IxyDj}{\mathcal{I}({\sysx\sysy}\! : \! {\sysj})}
\newcommand{\IxjDy}{\mathcal{I}({\sysx\sysj}\! : \! {\sysy})}
\newcommand{\IyjDx}{\mathcal{I}({\sysy\sysj}\! : \! {\sysx})}
\newcommand{\IxDy}{\mathcal{I}({\sysx}\! : \! {\sysy})}
\newcommand{\IxDj}{\mathcal{I}({\sysx}\! : \! {\sysj})}
\newcommand{\IyDj}{\mathcal{I}({\sysy}\! : \! {\sysj})}

\newcommand{\ExDy}{\mathcal{E}({\sysx}\! : \! {\sysy})}
\newcommand{\ExDj}{\mathcal{E}({\sysx}\! : \! {\sysj})}
\newcommand{\EyDj}{\mathcal{E}({\sysy}\! : \! {\sysj})}

\begin{document}

\preprint{APS/123-QED}

\title{Multipartite quantum correlations in a two-mode Dicke model}% Force lxine breaks with \\

\author{Rodolfo R. Soldati}
\email{rsoldati@usp.br}
\affiliation{Instituto de F\'isica da Universidade de S\~ao Paulo,  05314-970 S\~ao Paulo, Brazil.}
\affiliation{Institute for Theoretical Physics I, University of Stuttgart, D-70550 Stuttgart, Germany}
\author{Mark T. Mitchison}
\email{markTmitchison@gmail.com}
\affiliation{School of Physics, Trinity College Dublin, College Green, Dublin 2, Ireland}
\author{Gabriel T. Landi}
\email{gtlandi@gmail.com}
\affiliation{Instituto de F\'isica da Universidade de S\~ao Paulo,  05314-970 S\~ao Paulo, Brazil.}
\date{\today}

\begin{abstract}
We analyze multipartite  correlations in a generalized Dicke model involving two optical modes interacting with  an ensemble of two-level atoms.
In particular, we examine correlations beyond the standard bipartite entanglement and derive exact results in the thermodynamic limit. 
The model presents two superradiant phases involving the spontaneous breaking of either a \(\mathbb{Z}_2\) or U(1) symmetry.
The latter is characterized by the emergence of a Goldstone excitation, found to significantly affect the correlation profiles.
Focusing on the correlations between macroscopic subsystems, we analyze both the mutual information as well as the entanglement of formation for all possible bipartitions among the optical and matter degrees of freedom. It is found that while each mode entangles with the atoms, the bipartite entanglement between the modes is zero, and they share only classical correlations and quantum discord.
We also study the monogamy of multipartite entanglement and show that there exists genuine tripartite entanglement, i.e.~quantum correlations that the atoms share with the two modes but that are not shared with them individually, only in the vicinity of the critical lines. Our results elucidate the intricate correlation structures underlying superradiant phase transitions in multimode systems.

\end{abstract}

\maketitle

\section{Introduction}

Collective behavior in the quantum regime is directly associated with the emergence of complex correlation patterns among the individual particles~\cite{Osborne2001,*Osborne2002,Osterloh2002}.
Understanding how these correlations unfold is a crucial task in quantum many-body physics. 
For instance, the existence of an area or volume law for entanglement dictates what are the relevant corners of Hilbert space~\cite{Verstraete2008,Eisert2010}, and is crucial in characterizing the propagation of excitations in quantum chains~\cite{Calabrese2006}. An especially interesting question concerns the behavior of entanglement close to a quantum phase transition (QPT).
Historically, the first studies of this problem focused on the entanglement between one particle and the remainder of the system~\cite{Osborne2001}, or between pairs of particles~\cite{Osterloh2002,Arnesen2001}.
Focus subsequently shifted to the bipartite entanglement between two macroscopic parts of the system, which shows non-trivial scaling with subsystem size in the vicinity of the critical point~\cite{Vidal2003,Calabrese2005,Calabrese2005a,Wu2004a,Verstraete2004a,Wilms2011}.
These pioneering papers led to a 
surge of interest in entanglement in many-particle systems, especially in archetypal models such as one-dimensional spin chains (for a review, see Ref.~\cite{Amico2008}).
Experiments were also developed to directly assess bipartite correlations, using for instance interferometric techniques in ultra-cold atoms~\cite{Islam2015}, or tensor network tomography~\cite{Lanyon2017} in trapped ions.

A comprehensive understanding of the role of information in collective phenomena, however, requires one to move beyond the bipartite scenario by considering measures of genuine multipartite correlations~\cite{Rossini2021, Adesso2016}.
These quantities allow one to distinguish between fundamentally different types of correlation structures. For example, W and GHZ states of three qubits share the same type of bipartite entanglement, but entirely differ in their tripartite correlations, a property which can even be used to characterize two distinct classes of quantum states~\cite{Horodecki2009}. The situation becomes even more complex in genuine many-particle systems like spin chains~\cite{Wang2002,Santos2003,Bruss2005,Guehne2005}, due to the multiple different ways that subsystems can be correlated. The full hierarchy of multipartite correlations was recently studied in the Lipkin-Meshkov-Glick spin model~\cite{Lourenco2019}, as well as in the dynamics of superradiant light emission~\cite{Calegari2020}. However, most previous studies of multipartite entanglement in critical phenomena have focused on measures of the total entanglement shared among all constituent parts, e.g.~individual spins~\cite{DeOliveira2006,Wei2005,Hofmann2014,Motakhab2010,Pezze2017,Gabbrielli2018,Vidal2006}. It is therefore natural to also ask how multipartite entanglement near a QPT is shared between macroscopic or collective degrees of freedom.

An analytically tractable yet experimentally relevant setting in which to address this question is the superradiant phase transition induced by collective light-matter interactions, as described by the Dicke model~\cite{Wang1973,Hepp1973,Kirton2019}. Here, an optical field coupled to an atomic ensemble becomes occupied by a diverging number of photons when the coupling strength exceeds a critical value. \cite{Nagy2012,Wang2014}. Since the observation of this phase transition in a series of seminal experiments~\cite{Baumann2010,Baumann2011}, 
variants of the critical Dicke model involving multiple optical modes have been proposed~\cite{Fan2014a,Lang2017, Moodie2018, Chiacchio2018} and experimentally studied~\cite{Leonard2017,Leonard2017a}. 
This introduces rich new phenomenology in which QPTs arise from the interactions between a handful of collective modes instead of many small subsystems as in spin chains, for example.
The addition of extra modes can also enlarge the group of spontaneously broken symmetries~\cite{Fan2014a,Chiacchio2018}, leading to characteristic Goldstone and Higgs excitations~\cite{Lang2017,Leonard2017a}. 
However, while the standard Dicke model has been shown to exhibit diverging bipartite atom-field entanglement at the critical point~\cite{Lambert2004,*Lambert2005, Bakemeier2012,Vidal2005,Vidal2006}, little is known about the behavior of correlations in the multimode case.

To bridge this gap, we consider in this work a generalized Dicke model involving two optical modes interacting simultaneously with a large number of two-level atoms. This model was introduced in Ref.~\cite{Fan2014a} and motivated the experiments in Ref.~\cite{Leonard2017,Leonard2017a}. It boasts a rich ground-state phase diagram involving both first- and second-order superradiant QPTs, as well as a Goldstone mode that emerges along a critical line where a continuous rotation symmetry is broken, as we discuss in Sec.~\ref{sec:model}. Importantly, the system comprises three physically distinguished and effectively macroscopic subsystems: the atomic ensemble and the two optical modes (each one being able to support an arbitrarily large number of photons). Using tools from Gaussian continuous-variable quantum information theory, in Sec.~\ref{sec:correlations} we provide a thorough assessment of multipartite  correlations within all possible partitions of the system, among these three collective degrees of freedom.
We find that while the optical modes entangle with the atoms, bipartite correlations between the two modes are purely classical. Moreover, we find genuine tripartite entanglement emerging in the critical region. Interestingly, the critical behavior of bipartite entanglement depends on both the order of the transition and the partition considered, while the genuine tripartite entanglement is insensitive to the nature of the critical point. We discuss our results and conclude in Sec.~\ref{sec:discussion}. 
Units where \(\hbar=1\) are used throughout.

\section{Two-mode Dicke model with enlarged symmetry}
\label{sec:model}

\begin{figure}[!t]
\centering
    \includegraphics[width=\columnwidth]{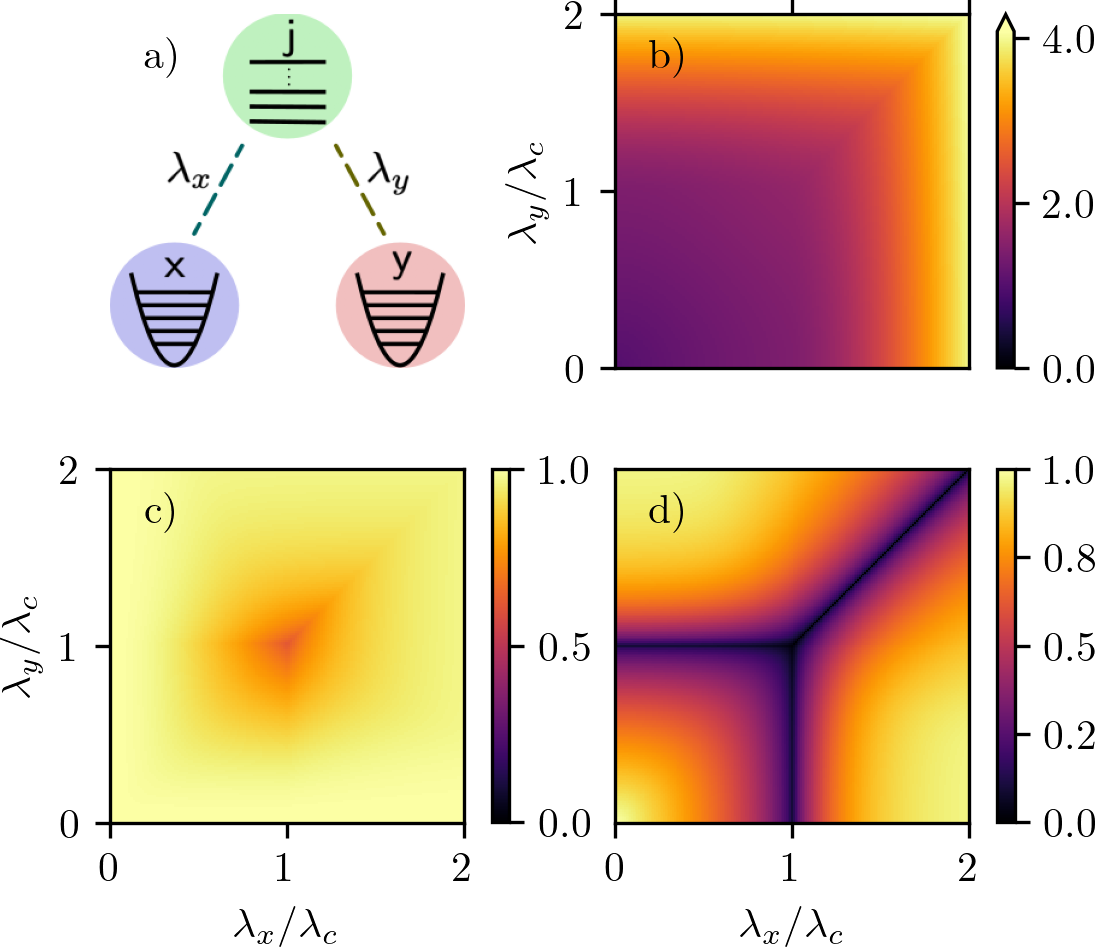}
    \caption{
    (a) Schematics of the two-mode Dicke model in Eq.~\eqref{eq:dicke-ham}. 
    (b, c, d) Energy gaps \(\nu_1\), \(\nu_2\) and \(\nu_3\), respectively, of excitation modes, showing the emergence of a Goldstone (gapless) mode, where \(\nu_3 = 0\), along the line \(\lx=\ly>\lc\). The minimum for (b) is \(\nu_1 = 1\) and for (c) is \(\nu_2 \approx 0.6\). {Note that in (b) the color scale covers a larger range than in (c) and (d).}
    }\label{fig:ordpar}
\end{figure}

We consider a system comprising two bosonic modes, with annihilation operators \(a_x\), \(a_y\), {and respective frequencies \(\omega_x\) and \(\omega_y\)}. Each mode interacts with an ensemble of \(N\) two-level atoms of frequency \(\omega_0\), which are described by collective spin operators \(J_x,J_y,J_z\), of total spin \(j = N/2\) (Fig.~\ref{fig:ordpar}(a)). 
The Hamiltonian is taken to be of the form
\begin{multline} \label{eq:dicke-ham}
    \ham = {\omega_x \ad_x a_x + \omega_y \ad_y a_y}
    + \omega_0 J_z \\
    + \frac{ \lx }{ \sqrt{ 2 j } } \left( \ad_x + a_x \right) J_x
    + \frac{ \ly }{ \sqrt{ 2 j } } \left( \ad_y + a_y \right) J_y ,
\end{multline}
where \(\lx,\ly \geqslant 0\) are the coupling strengths.
Other types of couplings are discussed in appendix~\ref{app:svd}.
In the thermodynamic limit, \(j \to \infty\), this model undergoes a superradiant QPT when {\(\operatorname{max}(\lx^2 / \omega_x, \ly^2 / \omega_y) > \omega_0\), which represents the spontaneous breaking of a \(\mathbb{Z}_2\).  
However, if \(\lx = \ly \) and \(\omega_x = \omega_y\)}, this is enlarged to a continuous U(1) symmetry, associated to the unitary \(\eu^{i \phi L_z}\). Here, \(\phi\) is an arbitrary parameter and \(L_z  = J_z + i(a_y^\dagger a_x - a_x^\dagger a_y)= J_z + q_x p_y - q_y p_x\) can be interpreted as the total angular momentum (spin + orbital), with  \(q_i = (a_i + a_i^\dagger)/\sqrt{2}\) and \(p_i = \iu(a_i^\dagger - a_i)/\sqrt{2}\) the position and momentum quadratures for mode \(i=x,y\). The breaking of this U(1) symmetry causes the appearance of a gapless Goldstone mode, i.e., a mode with zero frequency \cite{Baumann2010,Baumann2011,Leonard2017,Leonard2017a,Morales2019}.

The Hamiltonian~\eqref{eq:dicke-ham} can be diagonalized in the thermodynamic limit using a variety of methods~\cite{Wang1973,Hepp1973,Emary2003,Fan2014a,Kirton2019}. Here we use an approach similar to Ref.~\cite{Goes2020a}, which we have found particularly convenient.  
It consists of first determining a classical ground state, and then introducing the quantum fluctuations on top of it. In the thermodynamic limit, this procedure becomes exact.

\subsection{Classical ground state}
\label{sec:classical_ground_state}

To determine the classical ground state, we first replace all operators with \(c\)-numbers, \(a_i \mapsto \sqrt{2j}~\alpha_i\) and \((J_x,J_y,J_z) \mapsto j (\sin\theta\cos\phi, \sin\theta\sin\phi,\cos\theta)\).
This is tantamount to assuming that the ground state (GS) is a product of bosonic coherent states for the quadratures~\cite{Cahill1969} and spin coherent states for the macrospin~\cite{Radcliffe1971}.
Plugging this in~\eqref{eq:dicke-ham} yields the classical energy landscape \(H \to j E\), with 
\begin{equation} \label{eq:classical-gs}
    E = {\frac{\omega_x}{2} \alpha_x^2 + \frac{\omega_y}{2} \alpha_y^2} + \omega_0 \cos\theta + (\lambda_x \alpha_x \cos\phi + \lambda_y \alpha_y \sin\phi) \sin\theta.
\end{equation}
We now minimize this with respect to  \(\alpha_x, \alpha_y,\theta,\phi\). 
{Motivated by the appearance of an enlarged symmetry group, we henceforth assume \(\omega_x = \omega_y = \omega\), which leads to a QPT when \(\max(\lambda_x, \lambda_y)\) crosses the critical point \(\lambda_c = \sqrt{\omega\omega_0}\).
}
In the calculations that follow, we assume  \(\lambda_x > \lambda_y\); 
the case when \(\ly > \lx\) is treated similarly, by replacing \(x\leftrightarrow y\). 
Moreover, the case \(\lx \to \ly\) is always handled as a limit.

Under these assumptions, the minima always occur at \(\alpha_y = \phi = 0\). When \(\lx \leq \lc\), the minimum is found at
\begin{equation}
    \theta = \pi, \qquad \alpha_x = 0.
\end{equation}
This corresponds to the normal phase, with ground-state energy \(E_{\rm GS} = -\omega_0\). 
If \(\lx > \lc\), this solution becomes a maximum and a new minimum becomes available, at 
\begin{equation}
    \cos\theta = -\lc^2/\lx^2, 
    \qquad  \alpha_x = -(\lx / \omega) (1-\lc^4/\lx^4)^{1/2}.
\end{equation}
This represents the superradiant phase, with ground-state energy \(E_{\rm GS} =- (\lambda_x^4 + \lambda_c^4)/(2 \lambda_x^2 \omega)\).

The order of the QPTs separating these different phases can be inferred from the behavior of the ground-state energy across the critical point. The normal-superradiant transition is of second order because \(\partial^2 E_{\rm GS}/\partial \lambda_x^2\) is discontinuous across the critical line \(\lambda_x=\lambda_c\), while \(\partial E_{\rm GS}/\partial \lambda_x\) is continuous. Conversely, the transition between the two superradiant phases is of first order because \(\partial E_{\rm GS}/\partial \lx\) is discontinuous across the Goldstone line \(\lx=\ly>\lc\)~\cite{Fan2014a}.

\subsection{Quantum fluctuations}

Next, we introduce quantum fluctuations. 
To do that, we first change frames to the classical GS by applying a unitary 
\begin{equation}\label{eq:tham_def}
    \tham = U^\dagger H U, \qquad U = \eu^{-i \phi J_z} \eu^{-i \theta J_y} D_x(\alpha_x) D_y(\alpha_y), 
\end{equation}
using the same values of \(\alpha_x, \alpha_y,\theta,\phi\) obtained above. 
Here \(D_i(\alpha) = \eu^{\sqrt{j} \alpha (\ad_i - a_i)}\) are bosonic displacement operators. 
In this new frame, we then perform  a Holstein--Primakoff expansion~\cite{Holstein1940a} of the spin operators, by introducing \(J_x = \sqrt{j} Q\), \(J_y = \sqrt{j} P\) and \(J_z = j - (Q^2 + P^2)/2\), where \(Q\), \(P\) are canonical quadratures satisfying \([Q,P] = \iu\). 
This allows us to write \(\tham\) as a series expansion in powers of \(1/j\):
\begin{equation}\label{eq:tham}
    \tham = j E_{\rm GS} + \tham_2 + \mathcal{O}(j^{-1/2}), 
\end{equation}
where \(\tham_2\) is a \(j\)-independent Hamiltonian which is quadratic in the relevant operators. 
The  contribution of order \(j^{1/2}\) vanishes because of our choice of frame (i.e., our choice of \(\alpha_x,\alpha_y, \theta, \phi\)). Moreover, all others terms are at least of order \(j^{-1/2}\), thus also vanishing in the thermodynamic limit.

The actual form of \(\tham_2\) depends on which phase the system is in.
In terms of the mode quadratures \(q_i = (a_i + a_i^\dagger)/\sqrt{2}\) and \(p_i = \iu(a_i^\dagger - a_i)/\sqrt{2}\), one finds that in the normal phase 
\begin{equation} \label{eq:np}
    \tham_2^{\mathsf{np}} = \frac{\omega}{2}\sum_{i=x,y} (q_i^2 + p_i^2)
    + \frac{\lc^2}{2 \omega} (Q^2 + P^2) 
    + \lx Q q_x + \ly P q_y, 
\end{equation}
while in the superradiant phase 
\begin{equation} \label{eq:sp}
    \tham_2^{\mathsf{sp}} =  \frac{\omega}{2}\sum_{i=x,y} (q_i^2 + p_i^2)
    + \frac{\lx^2}{2 \omega} (Q^2 + P^2) 
    - \frac{\lc^2}{\lx} Q q_x + \ly P q_y .
\end{equation}
Both Hamiltonians are of the form 
\begin{equation}\label{eq:HK}
    \tham_2 = \frac{1}{2}\bm{r}^\trp K \bm{r},
\end{equation}
where \(\bm{r} = (q_x,p_x,q_y,p_y,Q,P)^\trp\) and \(K\) is a symmetric, positive-definite \(6\times 6\) matrix. 
The explicit form of \(K\) for the Hamiltonians~\eqref{eq:np} and~\eqref{eq:sp} are shown in {appendix~\ref{app:symp} in full generality}.
According to Williamson's theorem~\cite{Serafini2017}, 
it is always possible to find a symplectic matrix \(M\) such that 
\begin{equation}\label{eq:williamson}
    K = M V M^\trp, \qquad 
    V = \operatorname{diag}(\nu_1,\nu_1,\nu_2,\nu_2,\nu_3,\nu_3),
\end{equation}
where \(\nu_i \geqslant 0\) are the symplectic eigenvalues. 
A symplectic matrix is one which satisfies \(M \Omega M^\trp = \Omega\), where 
\begin{equation}
    \Omega = \bigoplus_{n=1}^3 \begin{pmatrix} 0 & 1 \\ -1 & 0 \end{pmatrix}
\end{equation}
is the symplectic form.
In our case, \(V\) and \(M\) must in general be computed numerically, which can be done using the algorithm in~\cite{Pirandola2009}.

Transforming to a new set of quadratures \(\bm{r}' = M^\trp \bm{r} = (q_1',p_1',q_2',p_2',q_3',p_3')^\trp\) (the ``normal modes''), 
finally puts \(\tham\) in diagonal form 
\begin{equation}\label{eq:tham_final}
    \tham = j E_{\rm GS} + \frac{1}{2} \sum\limits_{i=1}^3 \nu_i (q_i'^2 + p_i'^2) + \mathcal{O}(j^{-1/2}). 
\end{equation}
This establishes the nature of the quantum excitations that exist on top of the classical GS.
As can be seen, the excitations are bosonic, with equally spaced levels of gaps \(\nu_i\).
The dependence of the gaps \(\nu_1\), \(\nu_2\) and \(\nu_3\) are shown in Fig.~\ref{fig:ordpar}. 
Two of these modes, labeled \(\nu_{1,2}\), are always gapped (Fig.~\ref{fig:ordpar}(b) and (c)).
Conversely, the third mode \(\nu_3\) becomes gapless along the line \(\lx = \ly \geqslant \lc\) (Fig.~\ref{fig:ordpar}(d)).
This therefore represents the Goldstone mode. 
Since \(\bm{r}'\) is a linear combination of the original modes, the Goldstone mode is a collective excitation of the coupled light-matter system. 

\subsection{Quantum ground state}

To finish, we write down the full ground state, including the quantum fluctuations. 
Let \(\ket{0}\) be the state annihilated by the lowering operators associated with the original quadratures \(\bm{r}\).
The ground-state of \(\tham\) will be instead the vacuum \(|0'\rangle\) of the normal modes \(\bm{r}'\). 
The two are related via 
\(|0'\rangle = \exp( \frac{\iu}{2} \bm{r}^\trp G \bm{r} ) \ket{0}\), where \(G\) is a matrix associated to the eigenvectors \(M\) through \(M = \eu^{\Omega G}\)~\cite{Serafini2017}. 
The matrix \(G\) also relates \(\bm{r}\) and \(\bm{r}'\) according to 
\begin{equation}\label{eq:G_relation}
    e^{\frac{i}{2} \bm{r}^\trp G \bm{r} } \bm{r} e^{-\frac{i}{2} \bm{r}^\trp G \bm{r} } = e^{\Omega G} \bm{r} = M^\trp \bm{r} = \bm{r}'.
\end{equation}

Finally, we must also include the unitary \(U\) in Eq.~\eqref{eq:tham_def}. Note that, within a Holstein--Primakoff picture, the spin rotation \(\eu^{-i \theta J_y}\) also becomes a bosonic displacement operator. 
The ground state is therefore
\begin{equation}\label{eq:GS}
    \ket{\psi_\mathrm{GS}} = U \exp( \frac{\iu}{2} \bm{r}^\trp G \bm{r} ) \ket{0}, 
\end{equation}
The GS is thus a displaced squeezed state in the six quadratures \(\bm{r}\). 
The displacement is associated to the classical ground state, and is determined by the classical values \(\alpha_x, \alpha_y, \theta, \phi\).
The squeezing, on the other hand, is associated with the symplectic matrix \(M\) that diagonalizes \(K\).
This includes both single-mode squeezing, as well as two- (and multi-) mode squeezing among the quadratures. 
Hence, \(M\) is the matrix responsible for quantifying the correlations among different subsystems.

At the Goldstone line \(\lambda_x = \lambda_y > \lambda_c\), an arbitrariness emerges in the choice of ground-state, associated to the U(1) symmetry. 
As discussed in Sec.~\ref{sec:classical_ground_state}, we avoid this  by always taking \(\lambda_x \neq \lambda_y\), and interpreting the Goldstone line only as a  limit \(\lambda_x \to \lambda_y\). 
This is important because the correlations usually diverge on this line. Hence, in order to be able to meaningfully treat them, we must always consider that we are not exactly on the critical line, but in its vicinity.

In the thermodynamic limit, the Hamiltonian~\eqref{eq:tham} becomes quadratic. Hence, the ground state~\eqref{eq:GS} is Gaussian and thus fully characterized by the covariance matrix \(C_{\sf xyj}\) with entries
\begin{equation}\label{Cxyj}
    (C_{\sf xyj})_{k\ell} = \frac{1}{2} \mel*{\psi_{\rm GS}}{
        \acomm*{ r_k - \ev*{r_k} }{r_\ell - \ev*{r_\ell} } }{\psi_{\rm GS}},
\end{equation}
where \(\acomm{\ }{\ }\) is the anti-commutator.
Using Eq.~\eqref{eq:G_relation} one may conveniently rewrite this as 
\begin{equation} \label{eq:covariance}
    \cov_{\sf xyj} = \frac{1}{2} (M M^\trp)^{-1},
\end{equation}
Reduced states for any part of the system can be readily computed from this by simply deleting the entries of \(C_{\sf xyj}\) corresponding to the remainder. This includes the bipartite states \(C_{\sf xy}\), \(C_{\sf xj}\) and \(C_{\sf yj}\), as well as the single-party states \(C_{\sf x}\), \(C_{\sf y}\) and \(C_{\sf j}\).

{
We remark that at
the broken symmetry phase, where the ground state is degenerate, one could also use Eq.~\eqref{Cxyj} to construct the covariance matrix, for either superpositions or incoherent mixtures of ground-states.
However, mixtures of Gaussian states are not Gaussian.
Furthermore, away from the thermodynamic limit, at finite \(j\), the true ground state is also not expected to be Gaussian.
Either way, the underlying state will no longer be fully characterized solely by \(C_{\sf xyj}\).
Our results for the correlation profiles presented below therefore strictly hold only in the thermodynamic limit, as they often rely on tools specific for Gaussian states \cite{DeChiara2018}.
}

\section{Behavior of correlations around the critical point}
\label{sec:correlations}

We now proceed to analyze the correlation profiles in the ground state~\eqref{eq:GS}.
In terms of collective degrees of freedom, our system is effectively tripartite and we refer to each part as \(\sysx\), \(\sysy\) and \(\sysj\), respectively.  
The standard approach consists of analyzing correlations between the bipartitions \((\sysx\sysy,\sysj)\), \((\sysx\sysj,\sysy)\) and \((\sysy\sysj,\sysx)\). The latter becomes equivalent to \((\sysx\sysj,\sysy)\) under the exchange \(\lambda_x\to\lambda_y\), so there are effectively only two independent bipartitions to analyze. 
Since the global state of \(\sysx\sysy\sysj\) is pure, the standard measure of correlation is simply the entanglement entropy, i.e., the entropy of the reduced states. 

These correlations, however, fail to capture how the individual parties are correlated with each other. 
Thus, in addition, we will also study the correlations between \((\sysx,\sysy)\), \((\sysx,\sysj)\) and \((\sysy,\sysj)\) (which, again, is equivalent to \((\sysx,\sysj)\)). 
The reduced state of any two parties, however, such as \(\rho_{\sf xy} = \tr_{\sf j} |\psi_{\rm GS}\rangle\langle \psi_{\rm GS}|\), is mixed. Hence, the entropy is no longer a faithful quantifier of correlations.
This also introduces a fundamental distinction between classical correlations, quantum discord and quantum entanglement. 
Moreover, a conceptual difficulty arises, in that computing the actual entanglement (e.g.~through the relative entropy of entanglement~\cite{Horodecki2009}) becomes extraordinarily difficult.
We address these issues in two ways. First, we study the mutual information, which captures the net amount of correlations present (quantum and classical). 
And second, we compare this with the entanglement of formation~\cite{Bennett1996}, which provides an upper bound on the entanglement present in the system. 

\subsection{Mutual Information}

The net correlations (classical and quantum) between two modes {\sf A} and {\sf B}, with density matrix \(\rho_{\sf AB}\), can be quantified by the mutual information, 
\begin{equation} \label{eq:mut-info}
    \mathcal{I}( \mathsf{A}{:}\mathsf{B} ) = S(\rho_\mathsf{A}) + S(\rho_\mathsf{B}) - S(\rho_\mathsf{AB}),
\end{equation}
where \(S(\rho) = - \tr\rho\ln\rho\) is the von Neumann entropy, and \(\rho_{\sf A} = \tr_{\sf B} \rho_{\sf AB}\) and \(\rho_{\sf B}= \tr_{\sf A} \rho_{\sf AB}\) are the reduced states of {\sf A} and {\sf B}. 
If \(\rho_{\sf AB}\) is pure, Schmidt's theorem allows this to be reduced to  \(\mathcal{I}( \mathsf{A}{:}\mathsf{B} ) = 2 S(\rho_{\sf A}) = 2 S(\rho_{\sf B})\). 
This will be the case for the partitions \((\sysx \sysy,\sysj)\) and \((\sysx\sysj,\sysy)\):
\begin{align}
\label{eq:Ixy_j}
    \IxyDj ={}& 2 S(\sysj) = 2 S(\sysx\sysy),
    \\[0.2cm]
\label{eq:Ixj_y}    
    \IxjDy ={}& 2 S({\sysy}) = 2 S({\sysx\sysj}),
\end{align}
where \(S(\sysj) = S(\rho_\sysj)\) and so on.
The MI in these two cases is thus twice the entanglement entropy. 
Conversely, for \(\IxDy\) and \(\IxDj\) one must use the full definition in~\eqref{eq:mut-info}. 
Since the global state is pure, however, \(S(\sysx\sysy) = S(\sysj)\) and \(S(\sysx\sysj) = S(\sysy)\). Hence, one can partially simplify the expressions to 
\begin{align}
\label{eq:Ix_y}
    \IxDy ={}&  S(\sysx) + S(\sysy) - S(\sysj),
    \\[0.2cm]
\label{eq:Ix_j}    
    \IxDj ={}&  S(\sysx) + S(\sysj) - S(\sysy).
\end{align}
The fact that the global state is pure also allows us to relate Eqs.~\eqref{eq:Ixy_j}--\eqref{eq:Ixj_y} with Eqs.~\eqref{eq:Ix_y}--\eqref{eq:Ix_j}:
\begin{align}
\label{eq:I_rel_xyj}
    \IxyDj ={}& \IxDj + \IyDj, \\[0.2cm]
    \IxjDy ={}& \IxDy + \IyDj. 
\label{eq:I_rel_xjy}    
\end{align}

So far, we have been assuming that \(S(\rho)\) refers to the von Neumann entropy.
However, when dealing with information-theoretic quantifiers of Gaussian variables, it is more convenient to use the R\'enyi-2 entropy
\begin{equation}
S_2(\rho) = - \ln \tr \rho^2.     
\end{equation}
There are several reasons for this. 
First, the R\'enyi entropy is directly associated to the purity \(\tr \rho^2\) and thus has a clear physical interpretation.
For instance, in the ultra-cold atom experiment of Ref.~\cite{Islam2015}, the correlations were studied using the R\'enyi-2 entropy.
Second, for practically all Gaussian states, the two entropies are virtually indistinguishable. 
Third, \(S_2\) satisfies the strong subadditivity inequality for Gaussian states~\cite{Adesso2012}, which means it perfectly qualifies as an information-theoretic measure.
Finally, and most importantly, using the R\'enyi-2 entropy one can find a closed formula for the entanglement of formation, as will be discussed in Sec.~\ref{sec:eof}. If the von Neumann entropy were used instead, it would require a highly non-trivial minimization. 
The choice of R\'enyi-2 entropy therefore allows us to directly compare the mutual information with the entanglement of formation. 
For these reasons, throughout this paper \(S(\rho)\) will always refer to the R\'enyi-2 entropy, and the subscript 2 will be henceforth omitted.
Given a Gaussian state with CM \(C\), the R\'enyi-2 entropy is given simply by~\cite{Adesso2012}
\begin{equation}
    S(C) = \frac{1}{2} \ln \det 2C.
\end{equation}

\begin{figure}
    \includegraphics[width=\columnwidth]{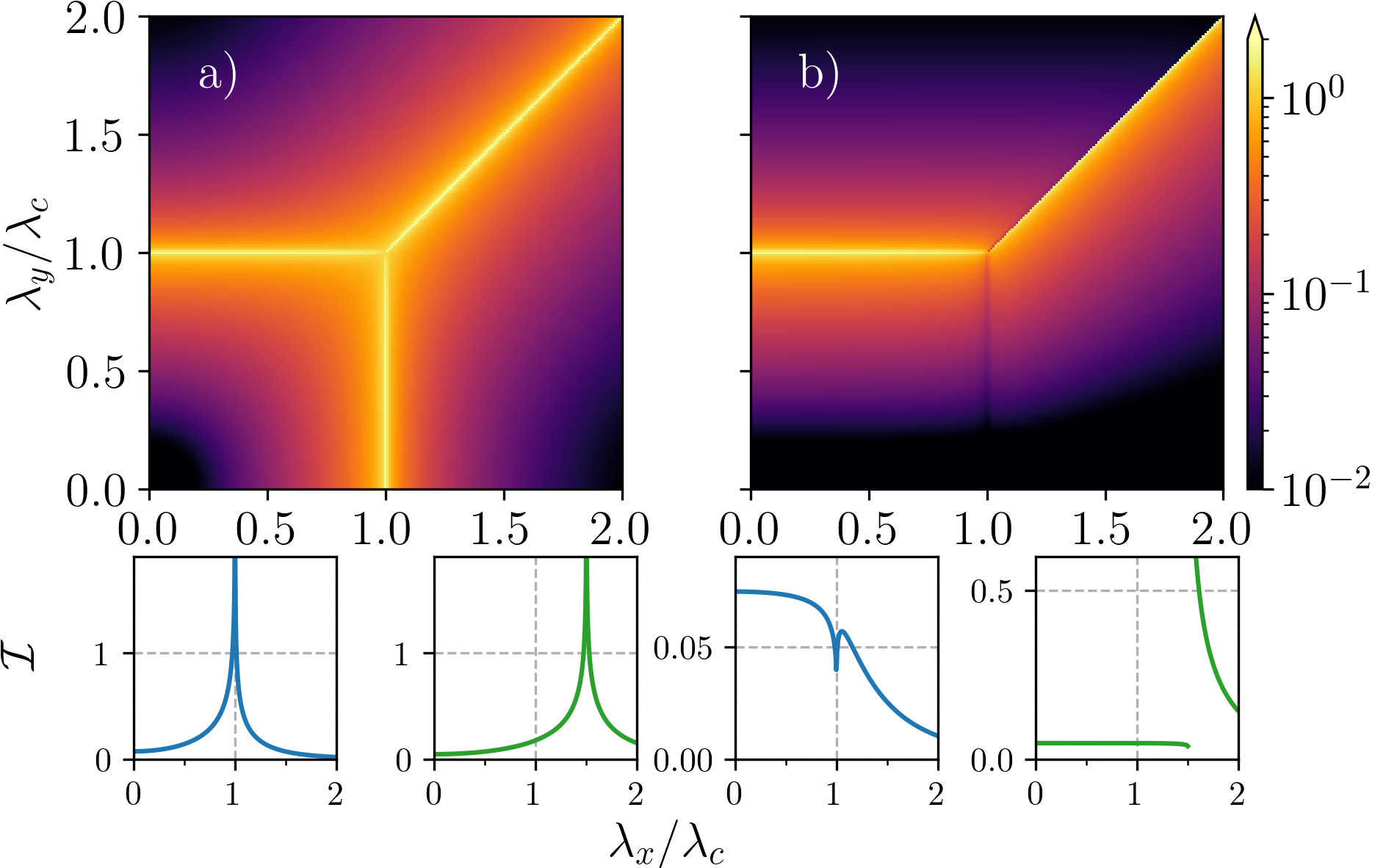}
    \caption{Mutual information for bipartitions of the global system, (a) \(\IxyDj\) and (b) \(\IxjDy\) in the \((\lambda_x,\lambda_y)\) plane. The plots at the bottom refer to slices at \(\lambda_y/\lambda_c = 0.5\) (blue) and 1.5 (green).
    The other bipartition \(\IyjDx\) is equivalent to \(\IxjDy\), but mirrored with respect to the line \(\lambda_x = \lambda_y\).
    Since the global state is pure, these quantities also represent twice the entanglement entropy between the parties, Eqs.~\eqref{eq:Ixy_j},~\eqref{eq:Ixj_y}.
    \label{fig:mutual-info}}
\end{figure}

In Fig.~\ref{fig:mutual-info} we present plots of \(\IxyDj\) and \(\IxjDy\) [Eqs.~\eqref{eq:Ixy_j} and~\eqref{eq:Ixj_y}] in the \((\lambda_x, \lambda_y)\) plane. 
The curves at the bottom correspond to slices at \(\lambda_y/\lambda_c = 0.5\) and \(\lambda_y/\lambda_c = 1.5\).
The remaining partition \(\IyjDx\) is identical to \(\IxjDy\), but mirrored with respect to the line \(\lambda_x = \lambda_y\).
In Fig.~\ref{fig:mutual-info}(a) we see that \(\IxyDj\) diverges at all critical lines. 
Moreover, it vanishes either when \(\lambda_{x,y}\to0\) or when \(\lambda_x \to \infty\) and \(\lambda_y \to \infty\). 
However, along the Goldstone line \(\lambda_x = \lambda_y\), it remains critical for arbitrarily strong interactions. 
Conversely, \(\IxjDy\) in Fig.~\ref{fig:mutual-info}(b) displays a much richer behavior: it has a kink if one crosses the critical line horizontally, a divergence if it is crossed vertically, and a discontinuity when the Goldstone line is crossed.
The behavior of \(\IxjDy\) therefore signals the order of the transition: it is discontinuous across the first-order QPT whereas its derivative is discontinuous across the second-order QPT, matching the expected behavior of bipartite entanglement near a quantum critical point~\cite{Wu2004a}. Note, however, that the same cannot be said about \(\IxyDj\) in Fig.~\ref{fig:mutual-info}(a).
This highlights some of the subtleties involved in the choice of partitions, which appear when dealing with models containing only a few collective modes. 

Next we turn to the MI between two of the three parties. 
The MI for \((\sysx,\sysj)\) and \((\sysx,\sysy)\) are shown in Figs.~\ref{fig:entangle}(a) and (b). 
We now find that \(\IxDj\) behaves similarly to the plot of \(\IxjDy\) [Fig.~\ref{fig:mutual-info}(b)], but with \(\lx\leftrightarrow \ly\).
Conversely, \(\IxDy\) is generally very small, being non-negligible only in the vicinity of the point \(\lambda_x, \lambda_y \sim \lambda_c\).
Both results are a consequence of the fact that \(\sysx\) and \(\sysy\) do not interact directly, but instead interact individually with \(\sysj\). 
This causes their individual correlations to be small, and thus \(\IxDj\sim \mathcal{I}(\sysx:\sysy\sysj)\) [Eq.~\eqref{eq:I_rel_xjy}], which is the mirror image of \(\IxjDy\) with respect to the line \(\lx=\ly\).
As illustrated by the slices in Figs.~\ref{fig:entangle},  \(\IxDj\) diverges as one crosses the critical line, and is either discontinuous or continuous depending on whether the transition is first or second order, as in Fig.~\ref{fig:mutual-info}(b). 
Note also that the additive property in Eq.~\eqref{eq:I_rel_xyj} explains why the total light-matter correlations \(\IxyDj\) shown in Fig.~\ref{fig:mutual-info}(a) lack any jump discontinuity across the first-order QPT.

\subsection{\label{sec:eof}Entanglement of Formation}

\begin{figure}[!t]
    \includegraphics[width=\columnwidth]{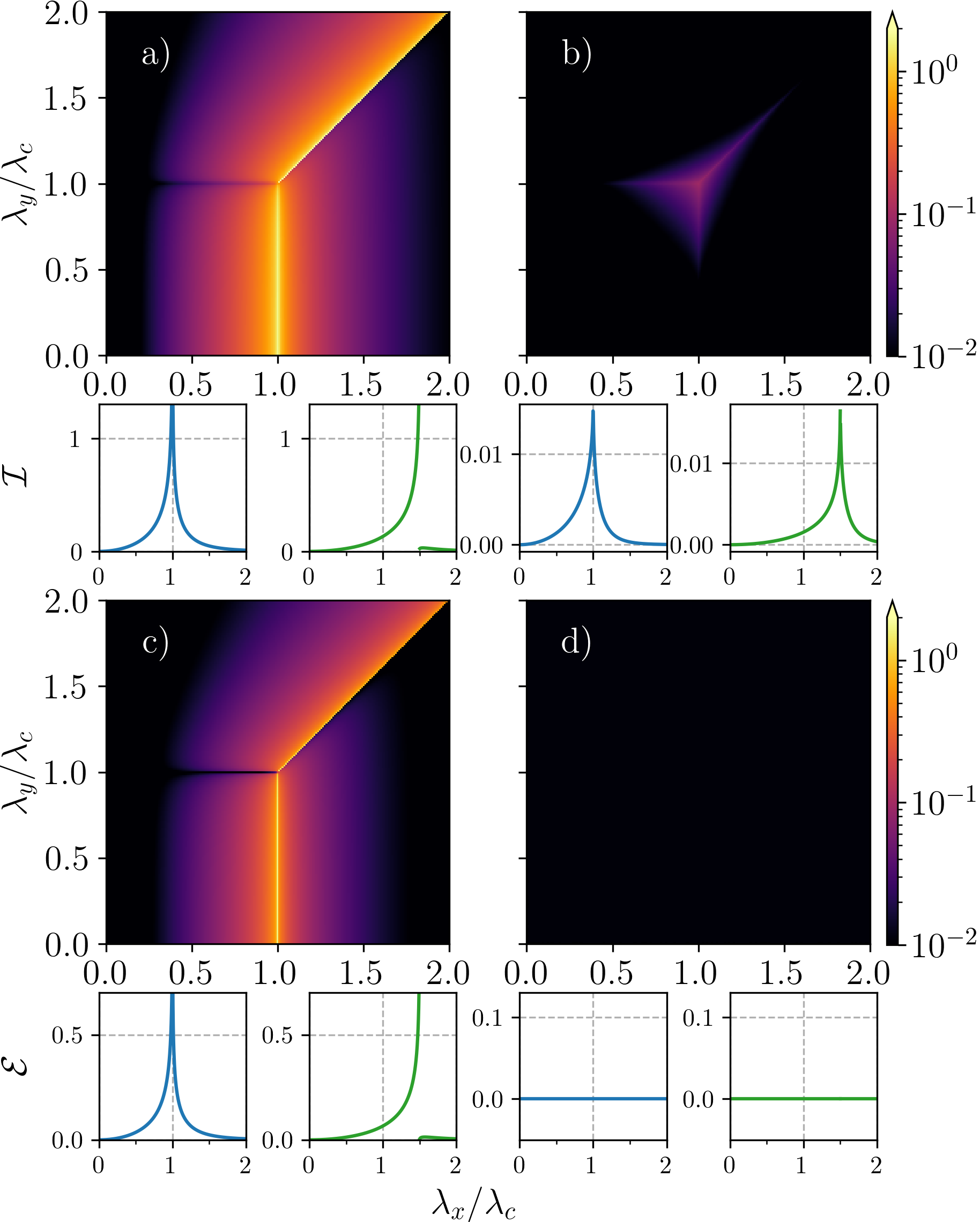}
    \caption{Mutual information between two parties, (a) \(\IxDj\) and (b) \(\IxDy\) [Eqs.~\eqref{eq:Ix_j} and~\eqref{eq:Ix_y}].
    (c), (d) Entanglement of formation~\eqref{eq:entanglement} for the same bipartitions. In the case of (x,y), image (d), the entanglement is identically zero.
    Other details are similar to Fig.~\ref{fig:mutual-info}.
    }\label{fig:entangle}
\end{figure}

Unlike the results in Fig.~\ref{fig:mutual-info}, those in Figs.~\ref{fig:entangle}(a) and (b) capture only the net correlations between the two parties, which is not necessarily entanglement. 
To further advance our analysis of the correlation profiles, we therefore now turn to an investigation of the degree of bipartite entanglement between the various collective degrees of freedom.

Since the reduced states are mixed, however, a direct calculation of an entanglement measure, e.g., the relative entropy of entanglement~\cite{Vedral1998}, becomes an extremely complicated task~\cite{Horodecki2009} that is in fact a long-standing open problem in quantum information theory~\cite{KruegerOpen}. Instead,  we focus here on bounding the entanglement via the entanglement of formation (EoF), introduced in~\cite{Bennett1996}. 
This is a type of convex roof extended measure, which quantifies the optimal entanglement cost to create a given state using only pure states together with local operations and classical communication.
Given a generic system \({\sf AB}\) with joint state \(\rho_{\sf AB}\),  the EoF is defined as 
\begin{equation} \label{eq:entanglement}
    \mathcal{E}( \mathsf{A}{:}\mathsf{B} ) = \inf_{p_i, \ket{\psi_i}} \sum_i p_i S( \ketbra{\psi_i}) ,
\end{equation}
where the minimization is taken over all possible decompositions  of the system state in the form \(\rho_{\sf AB} =\sum_i p_i |\psi_i\rangle\langle \psi_i| \).
For pure states, \(\mathcal{E}(\sf A{:}B) = \mathcal{I}(\sf A{:}B)/2\) reduces exactly to the entanglement entropy. 
Otherwise, \(\mathcal{E}(\sf A{:}B)\) provides an upper bound on the actual entanglement in the system.

In the case of Gaussian states, the sum can be replaced by an integral, and one may also use instead the R\'enyi-2 entropy. 
Restricting the minimization to Gaussian states yields a further upper bound for the EoF, which has the advantage of being expressible solely in terms of the elements of the covariance matrix. 
A closed-form expression for the EoF was developed in~\cite{Adesso2012, Adesso2014}, and is reported in Appendix~\ref{app:eof}.

In our problem, since the global state is pure, the quantities in Fig.~\ref{fig:mutual-info} already represent the EoF for the partitions \(({\sf xy,j})\) and \(({\sf xj,y})\). Instead, we therefore focus on \(\ExDj\) and \(\ExDy\), which are plotted in Figs.~\ref{fig:entangle}(c) and (d).
These can thus be directly compared with the corresponding mutual informations \(\IxDj\) and \(\IxDy\). 
As can be seen, \(\ExDj\) behaves similarly to the MI \(\IxDj\). 
In fact, from the slices below it becomes clear that the scale of \(\ExDj\) is roughly half that of \(\IxDj\).
This therefore shows that for these two parties, most of the correlations are actually made up from entanglement. 
In other words, most of the correlation between the optical modes and the macrospin \(\sysj\) are in the form of entanglement. This, as is expected, is maximized closed to the critical line. 
At \(\lambda_x \sim \lambda_c\) and small \(\lambda_y\), the entanglement \(\ExDj\) is large, while \(\EyDj\) would be small, and vice-versa. 
On the other hand, at the Goldstone line both \(\ExDj\) and \(\EyDj\) will be large. 

Next we turn to the correlations between the two optical modes, \(\IxDy\) and \(\ExDy\), Figs.~\ref{fig:entangle}(b) and (d). 
As can be seen, while \(\IxDy\) is generally small, \(\ExDy\) is \emph{identically zero}. 
This is again a consequence of the fact that \(\sysx\) and \(\sysy\) do not interact directly. 
Thus, even though \(\sysj\) is capable of mediating a finite (albeit small) mutual information between them (Fig.~\ref{fig:entangle}(b)), the entanglement is identically zero. 
Interestingly, even though the EoF is only an upper bound, the fact that it is identically zero suffices to show that the entanglement is itself identically zero. 
This leads to the important conclusion that the ground state~\eqref{eq:GS} is  separable. 
{Clearly, this conclusion is expected to change dramatically if there was any direct interaction between the two modes. Of course, this will depend on the strength of said interaction. A future investigation of the effect of such direct interactions would be of great interest, especially since they are now within reach of experiments as reported recently in Ref.~\cite{Morales2018}.
}

It is also worth mentioning that the reduced state of \(\sysx\sysy\) does not contain only classical correlations, but also contains quantum discord. This is a consequence of theorem proved in~\cite{Giorda2010}, which establishes that any two-mode squeezed state has non-zero discord.

\subsection{Tripartite entanglement}

In the present case, the EoF~\eqref{eq:entanglement} satisfies the monogamy property~\cite{Coffman2000a, Adesso2012}
\begin{equation} \label{eq:monogamy}
    \mathcal{E}({\sf i{;}j{:}k }) \equiv  \mathcal{E}({\sf i{:}jk} ) - \mathcal{E}({\sf i{:}j }) - \mathcal{E}({\sf i{:}k }) \geq 0 ,
\end{equation}
for any three parties \({\sf i, j, k}\). We note that while the EoF is not monogamous in general~\cite{Lancien2016}, Eq.~\eqref{eq:monogamy} does hold for the Rényi-2 EoF on Gaussian states~\cite{Adesso2012}. Therefore, a non-zero value of \(\mathcal{E}({\sf i{;}j{:}k })\) means that there exist correlations that \({\sf i}\) shares collectively with \({\sf jk}\), but which are not shared individually with \({\sf j}\) and \({\sf k}\).
Compare, for instance, with Eqs.~\eqref{eq:I_rel_xyj} and~\eqref{eq:I_rel_xjy}.
The mismatch \(\mathcal{E}({\sf i{;}j{:}k})\) therefore quantifies the genuine tripartite entanglement, as it describes correlations which reside on the global space of three parties, but are not present in the reduced space of the different pairs. 
Note that according to this definition \(\mathcal{E}({\sf i{;}j{:}k} )\) is not permutation invariant, but is defined with respect to the first index \({\sf i}\), if the global state is pure, as in our case,  \(\mathcal{E}({\sf i{:}jk}) = \mathcal{I}({\sf i{:}jk})/2 = S(\rho_{\sf i}) = S(\rho_{\sf jk})\), which partially simplifies the expression for \(\mathcal{E}({\sf i{;}j{:}k })\).

\begin{figure}
    \includegraphics[width=\columnwidth]{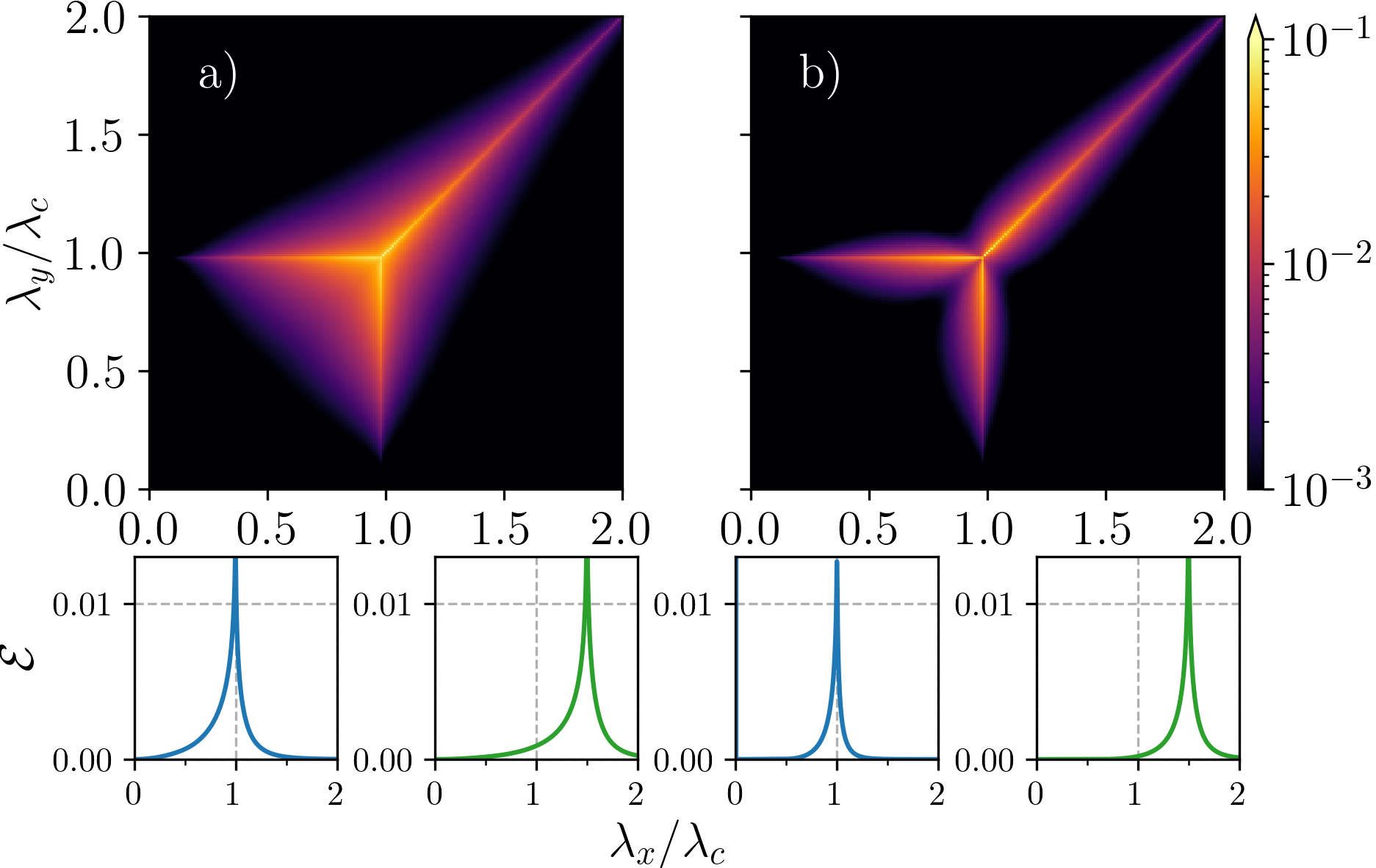}
    \caption{Tripartite entanglement (a) \(\mathcal{E}(\sysx{;}\sysy{:}\sysj)\) and (b) \(\mathcal{E}(\sysj{;}\sysy{:}\sysx)\), computed from Eq.~\eqref{eq:monogamy}.
    Other details are similar to Fig.~\ref{fig:mutual-info}.
    \label{fig:tri-entangle}
    }
\end{figure}

The tripartite entanglement measures \(\mathcal{E}(\sysx{;}\sysy{:}\sysj)\) and \(\mathcal{E}(\sysj{;}\sysy{:}\sysx)\) are presented in Fig.~\ref{fig:tri-entangle}.
We find that both quantities are significant only close to the critical lines. 
Concerning \(\mathcal{E}(\sysj{;}\sysy{:}\sysx)\)  in Fig.~\ref{fig:tri-entangle}(b), since \(\mathcal{E}(\sysx{:}\sysy)= 0\) (c.f.~Fig.~\ref{fig:entangle}(d)), we get
\begin{equation}
    \mathcal{E}(\sysj{;}\sysy{:}\sysx) = S(\sysx) - \mathcal{E}(\sysx{:}\sysj). 
\end{equation}
Conversely, for \(\mathcal{E}(\sysx{;}\sysy{:}\sysj)\) we get instead 
\begin{equation}\label{eq:exyj1}
    \mathcal{E}(\sysx{;}\sysy{:}\sysj) = S(\sysx\sysy) - \ExDj - \EyDj. 
\end{equation}
However, from the results of Fig.~\ref{fig:entangle}(b), we see that \(\sysx\sysy\) are only weakly correlated. As a rough approximation, we could thus take \(S(\sysx\sysy) \simeq S(\sysx) + S(\sysy)\), although this holds only outside the critical lines. 
In this case Eq.~\eqref{eq:exyj1} simplifies to 
\begin{align}
    \mathcal{E}(\sysx{;}\sysy{:}\sysj) \simeq{}& S(\sysx) - \ExDj + S(\sysy) - \EyDj \nonumber \\[0.2cm]
    ={}& \mathcal{E}(\sysj{;}\sysy{:}\sysx) + S(\sysy) - \EyDj. 
\end{align}
The quantity on the LHS is plotted in Fig.~\ref{fig:tri-entangle}(a), while \(\mathcal{E}(\sysj{;}\sysy{:}\sysx)\) is plotted in Fig.~\ref{fig:tri-entangle}(b).
This therefore explains why one quantity is much larger than the other. 
It is also interesting to note how both \(\mathcal{E}(\sysx{;}\sysy{:}\sysj)\) and \(\mathcal{E}(\sysj{;}\sysy{:}\sysx)\)  are insensitive to the order of transition, showing the same type of peaked behavior at either the blue or the green slices in Fig.~\ref{fig:tri-entangle}.

\section{Discussion}\label{sec:discussion}

We {have analyzed} the correlation profiles of a two-mode Dicke model undergoing both first- and second-order superradiant QPTs~\cite{Fan2014a}. Particular emphasis was given to the special Goldstone line, where the discrete \(\mathbb{Z}_2\) symmetry of the standard Dicke model is upgraded to a continuous U(1) symmetry, as studied in recent experiments~\cite{Leonard2017,Leonard2017a}. By focusing our attention on macroscopic degrees of freedom in the thermodynamic limit --- namely, the atomic ensemble and the many-photon state of the two optical modes --- we simplified the description of the system into an effective three-body problem. This allowed us to consider in detail both quantum and classical correlations among all three parties, while distinguishing between bipartite and genuinely multipartite correlations.

Our results indicate that each superradiant QPT is primarily driven by diverging bipartite entanglement, which is shared between the atomic ensemble and the particular optical mode that becomes macroscopically occupied upon traversing the critical point. Although this finding is broadly consistent with the entanglement characteristics of the standard (single-mode) Dicke QPT~\cite{Lambert2004}, it is nonetheless non-trivial because the state of the two relevant parties is far from pure in our case {(the global tripartite state is pure, but the reduced state of any pair of subsystems is mixed)}. The bipartite entanglement shows the expected singular behavior depending on whether the QPT is of first or second order~\cite{Wu2004a}. Conversely, the bipartite entanglement between the two optical modes is strictly zero across the entire phase diagram, despite their classical bipartite correlations peaking on the critical lines. This feature can be understood from the absence of any direct interaction between the two modes. Nevertheless, we found the existence of genuine multipartite entanglement between all three parties, which peaks in the critical region. Interestingly, the critical behavior of the genuine multipartite entanglement between these physically distinct, macroscopic subsystems does not discriminate between first- and second-order QPTs, in contrast to the critical entanglement shared among all particles of a homogeneous system~\cite{DeOliveira2006}.

To obtain these results we have derived full analytical solutions for the various information-theoretic quantities, which become exact in the thermodynamic limit. These solutions are expressed in terms of the ground-state covariance matrix, which can be experimentally accessed by measuring the corresponding mode quadratures and variances, e.g. by heterodyne detection~\cite{Brunelli2018}. 
{The use of Gaussian tools for characterizing the correlation profiles was crucial to obtain our results, especially for entanglement. 
For non-Gaussian states, such as mixtures of Gaussian states or the ground state at finite system size, a similar analysis would be considerably more difficult.}
Nevertheless, our methods can be applied to several other systems in which superradiant phase transitions of both first and second order have been discovered~\cite{Asshab2013,Baksic2014, Carmichael2015,Hwang2015,Hwang2016,Liu2017,Peng2019,Felicetti2020,Shapiro2020,Cai2021}. 
{Conversely, the correlation profiles in models for which the state is not Gaussian can in principle be measured using, for example, the interferometric methods used in Ref.~\cite{Islam2015} or a many-body entanglement witness, such as that used in Ref.~\cite{Schmied2016}.}

Finally, and quite interestingly, our methods are directly applicable to open quantum systems. In this case, quantum fluctuations compete with dissipation to produce novel phases~\cite{Nagy2011,Nagy2015,*Nagy2016,Klinder2015,Hwang2018}. 
{The impact of the open dynamics on the critical properties can be highly non-trivial, with effects such as structured noise spectra playing an important role~\cite{Nagy2015}. 
Concerning the mutual information, one would intuitively expect that dissipation would generally weaken it, although we cannot rule out scenarios where it may be enhanced.
The behavior of the entanglement, on the other hand, is more unpredictable. 
Due to the minimization appearing in Eq. (26), the entanglement can suddenly change from non-zero to identically zero. 
We thus cannot preclude the possibility that the entanglement profile in the \((\lambda_x,\lambda_y)\) plane would be dramatically altered by dissipation.
For these reasons,} a detailed analysis of our model in the presence of dissipation will be presented in future work.

\begin{acknowledgments}
The authors acknowledge fruitful discussions with M.-J. Hwang and G. Adesso. We also thank M. Huber for useful comments on the manuscript.
R.R.S. acknowledges the financial support from the Brazilian funding agency CNPq (141797/2019-3) and the DAAD research grant Bi-nationally Supervised Doctoral Degrees/Cotutelle, 2020/21 (57507869).
M.T.M. acknowledges funding from the European Research Council Starting Grant ODYSSEY (Grant Agreement No. 758403) and the EPSRC-SFI Joint Funding of Research project QuamNESS.
G.T.L. acknowledges the financial support from the S\~ao Paulo Research Foundation (FAPESP) and the Brazilian funding agency CNPq.
\end{acknowledgments}

\appendix

\section{{\label{app:svd}More general interactions}}

{
A more general Hamiltonian than \eqref{eq:dicke-ham} is
\begin{equation}\label{app_dicke_gen}
    \ham = \omega_x a_x^\dag a_x + \omega_y a_y^\dag a_y + \omega_0 J_z
    + \frac{1}{\sqrt{2j}} \bm{q}^\trp \Lambda \bm{J} ,
\end{equation}
where the matrix \(\Lambda\) couples the vector operators \(\bm{q} = (q_x, q_y)\) and \(\bm{J} = (J_x, J_y)\), and has the form
\begin{equation}
    \Lambda =
    \begin{pmatrix}
        \Lambda_{xx} & \Lambda_{xy} \\
        \Lambda_{yx} & \Lambda_{yy}
    \end{pmatrix} .
\end{equation}
Note that in general \(\Lambda\) does not have to be Hermitian.
The Hamiltonian~\eqref{app_dicke_gen} can be connected to that in Eq.~\eqref{eq:dicke-ham} by introducing the singular value decomposition of \(\Lambda\) 
\begin{equation}
    \Lambda = U \bm{\lambda} V^\dag ,
\end{equation}
where \(\bm{\lambda} = \operatorname{diag}( \lambda_x, \lambda_y )\) are the non-negative singular values, and \(U, V\) are unitary matrices. We now define the rotated set of operators \(\tilde{\bm{q}} = U^\dag \bm{q}\), 
\(\tilde{\bm{p}} = U^\dag \bm{p}\), 
and \(\tilde{\bm{J}} = V^\dag \bm{J}\),
\(\tilde{J}_z = J_z\),
which transforms Eq.~\eqref{app_dicke_gen} to a form exactly like~\eqref{eq:dicke-ham}, with new couplings \(\lambda_x,\lambda_y\) now given by the singular values (note also that this rotation does not affect the first line of Eq.~\eqref{eq:dicke-ham}).

This means that as far as the critical properties are concerned, the two Hamiltonians are therefore equivalent. 
In fact, when working with the general form~\eqref{app_dicke_gen}, one may verify that all critical properties can be cast in terms of the matrix 
\begin{equation}
    R = \Lambda^\trp \Omega^{-1} \Lambda =
    \begin{pmatrix}
        r_{xx} & r_{xy} \\
        r_{xy} & r_{yy}
    \end{pmatrix} ,
\end{equation}
which \(\Omega = {\rm Diag}(\omega_x, \omega_y)\). Note that \(R\) is symmetric and positive semidefinite by construction. 
For example, the minimum of the classical energy associated to Eq.~\eqref{app_dicke_gen} occurs when 
\begin{equation}
    \tan(2\phi) = \frac{2 r_{xy}}{r_{xx} - r_{yy}}. 
\end{equation}
And the angle \(\theta\) is given by \(\cos \theta = -\omega_0/r\), where 
\(r\)
is the largest eigenvalue of \(R\).
Finally, the coherent state variables are given by \((\alpha_x, \alpha_y) = -\Omega^{-1} \Lambda \Phi (1-\omega_0^2/r^2)^{1/2}\), where \(\Phi = (\cos\phi, \sin\phi)\). 
This superradiant phase will only exist provided \(r > \omega_0\), which determines the critical line of the model. 
If this is not the case, the only solution will be the normal phase \(\theta = 0\). 
For the case where \(\Lambda\) is diagonal, the matrix \(R\) reduces to \(R = {\rm diag}(\lambda_x^2/\omega_x,\lambda_y^2/\omega_y)\). 
Hence, in this case the critical line becomes 
\(r = \max(\lambda_x^2/\omega_x, \lambda_y^2/\omega_y) > \omega_0\), which is the result stated below Eq.~\eqref{eq:dicke-ham}. 
Despite this connection between Eqs.~\eqref{app_dicke_gen} and~\eqref{eq:dicke-ham}, it is not possible to make analogous predictions for the correlation profiles, since they depend sensitively on the partitions being used; that is, on which combinations of modes are the correlations being analyzed.}

\section{\label{app:symp}Symplectic diagonalization}

The Hamiltonians~\eqref{eq:np} and \eqref{eq:sp}, are quadratic and can thus be written in the form~\eqref{eq:HK}, with 

\begin{equation}
    K^\mathsf{np} =
    \begin{pmatrix}
     \omega_x & 0 & 0 & 0 & \lx & 0 \\
     0 & \omega_x & 0 & 0 & 0 & 0 \\
     0 & 0 & \omega_y & 0 & 0 & \ly \\
     0 & 0 & 0 & \omega_y & 0 & 0 \\
     \lx & 0 & 0 & 0 & \omega_0 & 0 \\
     0 & 0 & \ly & 0 & 0 & \omega_0\\
    \end{pmatrix},
\end{equation}
and 
\begin{equation}
    K^\mathsf{spx} =
    \begin{pmatrix}
     \omega_x & 0 & 0 & 0 & -\omega_0 \omega_x / \lx & 0 \\
     0 & \omega_x & 0 & 0 & 0 & 0 \\
     0 & 0 & \omega_y & 0 & 0 & \ly \\
     0 & 0 & 0 & \omega_y & 0 & 0 \\
     -\omega_0 \omega_x / \lx & 0 & 0 & 0 & \lx^2 / \omega_x & 0 \\
     0 & 0 & \ly & 0 & 0 & \lx^2 / \omega_x \\
    \end{pmatrix} \label{eq:qfnpsp}.
\end{equation}
As used throughout the main text, the latter refers to the case { \(\lambda_x\sqrt{\omega_y} > \lambda_y\sqrt{\omega_x}\). 
The case \(\lambda_y\sqrt{\omega_x} > \lambda_x \sqrt{\omega_y}\)} can be obtained by simply reversing the roles of \(x\) and \(y\). 
The corresponding matrix \(K\) would thus become
\begin{equation}
    K^\mathsf{spy} =
    \begin{pmatrix}
    \omega_x & 0 & 0 & 0 & 0 & \lx \\
    0 & \omega_x & 0 & 0 & 0 & 0 \\
    0 & 0 & \omega_y & 0 & \omega_0 \omega_y / \ly & 0 \\
    0 & 0 & 0 & \omega_y & 0 & 0 \\
    0 & 0 & \omega_0 \omega_y / \ly & 0 & \ly^2 / \omega_y & 0 \\
    \lx & 0 & 0 & 0 & 0 & \ly^2 / \omega_y
    \end{pmatrix} . \label{eq:qfspy}
\end{equation}
These matrices are the basis for computing the symplectic eigenvalues and eigenvectors, \(V\) and \(M\), in Eq.~\eqref{eq:williamson}. 
As mentioned in the main text, this can be done using the algorithm in~\cite{Pirandola2009}. 

\section{\label{app:eof}R\'enyi-2 Entanglement of Formation}

The minimization of the R\'enyi-2 entanglement of formation,  Eq.~\eqref{eq:entanglement}, when restricted Gaussian pure states, can be written in the form~\cite{Adesso2014} 
\begin{equation}\label{EoF_Gaussian}
    \mathcal{E}( \mathsf{A}{:}\mathsf{B} ) = \inf_{\substack{ \{\Gamma_\mathsf{AB}:~ 0\leq \Gamma_\mathsf{AB} \leq \cov_\mathsf{AB},\\ \det(2\Gamma_{\rm AB}) = 1 \}}} \frac{1}{2} \ln( \det 2 \Gamma_\mathsf{A} ) .
\end{equation}
Here \(C_{\rm AB}\) is the CM of the joint system AB, and the minimization is over the set of pure Gaussian states (\({\rm det}(2\Gamma_{\rm AB}) = 1\). The function being minimized, \(\frac{1}{2} \ln \text{det}(2\Gamma_A)\), is thus nothing but the R\'enyi-2 entropy of the reduced state \(\Gamma_{\rm A}\). 
When \(C_{\rm AB}\) is pure, this reduces to 
the standard entanglement entropy
\begin{equation} \label{eq:bi-entanglement}
    \mathcal{E}( \mathsf{A}{:}\mathsf{B} ) = \frac{1}{2} \ln\det 2\cov_\mathsf{A} = \frac{1}{2} \ln\det 2\cov_\mathsf{B} .
\end{equation}
The minimization in Eq.~\eqref{EoF_Gaussian} can be performed analytically. 

To accomplish this it is convenient, however, to first put the CM in standard form, via local unitaries. 
Since we are interested in the entanglement involving two or three modes, we discuss here the algorithm for putting a generic \emph{pure} 3-mode CM in standard form, following~\cite{Adesso2014}.
Such a standard form CM has the form 
\begin{equation} \label{eq:covstdform}
    2C =
    \begin{pmatrix}
        a_1 & 0 & c_{3}^{+} & 0 & c_{2}^{+} & 0 \\
        0 & a_1 & 0 & c_{3}^{-} & 0 & c_{2}^{-} \\
        c_{3}^{+} & 0 & a_2 & 0 & c_{1}^{+} & 0 \\
        0 & c_{3}^{-} & 0 & a_2 & 0 & c_{1}^{-} \\
        c_{2}^{+} & 0 & c_{1}^{+} & 0 & a_3 & 0 \\
        0 & c_{2}^{-} & 0 & c_{1}^{-} & 0 & a_3
    \end{pmatrix},
\end{equation}
where the factor of 2 is placed in order to match with the notation of~\cite{Adesso2014}.
Moreover, since the state is pure, the coefficients \(c_i^\pm\) are related to the \(a_i\) according to 
{
\begin{equation}
\begin{split}
    4 \sqrt{a_{j} a_{k}} c_{i}^{\pm} ={}& \sqrt{ \left(a_{i}-1\right)^{2}-(a_{j}-a_{k})^{2} } \\
    &\times \sqrt{ (a_{i}+1)^{2}-(a_{j}-a_{k})^{2} } \\
    &\pm \sqrt{ (a_{i}-1)^{2}-(a_{j}+a_{k})^{2} } \\
    &\times \sqrt{ (a_{i}+1)^{2}-(a_{j}+a_{k})^{2} }
\end{split}
\end{equation}
}
Any 3-mode CM can be put in this form by means of local unitaries (which do not affect any measures of correlations).

Given a generic 3-mode CM, to put it in standard form one first computes the symplectic diagonalization of each \(2\times 2\) diagonal block (corresponding to the reduced states of each mode). 
This results in a local symplectic matrix 
\(\mathcal{M} = M_{\sf A} \oplus M_{\sf B} \oplus M_{\sf C}\).
The resulting matrix \(\mathcal{M} C \mathcal{M}^\trp\) will either already in standard form (in which case nothing more is done), or will be in one of the following schematic forms, with each block representing a \(2 \times 2\) submatrix. Letting \(C\) be the covariance matrix of our system, then
\begin{equation}\label{appB_Mcal}
    \mathcal{M} \cov \mathcal{M}^\trp =
    \begin{pmatrix}
        \boxbslash &  \boxslash & \boxslash \\
        \boxslash & \boxbslash & \boxbslash \\
        \boxslash & \boxbslash & \boxbslash
    \end{pmatrix},
    \quad \text{or }\ 
     \begin{pmatrix}
        \boxbslash & \boxbslash & \boxslash \\
        \boxbslash & \boxbslash & \boxslash \\
        \boxslash & \boxslash & \boxbslash
    \end{pmatrix},
    \quad \text{or }\ 
    \begin{pmatrix}
        \boxbslash & \boxslash & \boxbslash \\
        \boxslash & \boxbslash & \boxslash \\
        \boxbslash & \boxslash & \boxbslash
    \end{pmatrix} .
\end{equation}
Here \(\boxbslash\) and \(\boxslash\) mean, respectively, a diagonal or anti-diagonal block.
Depending on which case one gets, the CM can be cast in final form by multiplying by a matrix \(\mathcal{T}\), such as 
\begin{equation}
    C_{\text{std. form}} = \mathcal{T} \mathcal{M} C \mathcal{M}^\trp \mathcal{T}^\trp.
\end{equation}
The matrix \(\mathcal{T}\) depends on the shape of the resulting matrix in~\eqref{appB_Mcal}, and can be either:
\begin{equation}
    \mathcal{T} = \begin{pmatrix}
        1 & 0 & 0 \\
        0 & \Omega_2 & 0 \\
        0 & 0 & \Omega_2
    \end{pmatrix},
    \quad \text{or }\ 
    \begin{pmatrix}
        \Omega_2 & 0 & 0 \\
        0 & \Omega_2 & 0 \\
        0 & 0 & 1
    \end{pmatrix},
    \quad \text{or }\ 
    \begin{pmatrix}
        1 & 0 & 0 \\ 
        0 & \Omega_2 & 0 \\
        0 & 0 & 1,
    \end{pmatrix},
\end{equation}
where all matrices are \(2\times 2\) blocks, and \(\Omega_2 = \left(\begin{smallmatrix} 0 & 1 \\ -1 & 0 \end{smallmatrix}\right)\) is the single-mode symplectic form.

Finally, with the CM in diagonal form, we can now use the analytical results of~\cite{Adesso2014}, who showed that the bipartite entanglement \(\mathcal{E}({\sf A}_i {:} {\sf A}_j)\) between any 2 of 3 modes in a CM of the form~\eqref{eq:covstdform} is 
\begin{equation} \label{eq:bi-entanglement-mixed}
    \mathcal{E}( \mathsf{A}_i{:}\mathsf{A}_j ) = \frac{1}{2} \ln g_k
\end{equation}
where \(\{i,j,k\}\) is a permutation of \(\{1,2,3\}\),

\begin{equation}
    g_k =
    \begin{dcases}
        1 &\qif a_k \geq \sqrt{a_i^2+a_j^2-1} \\
        \frac{\beta }{ 8 a_k^2 } &\qif \alpha_{ij} < a_k < \sqrt{a_i^2+a_j^2-1} \\
        \frac{(a_i^2-a_j^2)^2 }{ (a_k^2-1)^2 } &\qif a_k \leq \alpha_{ij},
    \end{dcases} 
\end{equation}
and
\begin{widetext}
\begin{subequations}
\begin{align}
    \alpha_{ij} ={}& \left( \frac{ 2 (a_{i}^2+a_{j}^2) + (a_{i}^2-a_{j}^2)^2 + |a_{i}^2-a_{j}^2| \sqrt{ (a_{i}^2-a_{j}^2)^2 + 8 (a_{i}^2+a_{j}^2) } }{ 2 (a_{i}^2+a_{j}^2) } \right)^{1/2} \\
    \beta ={}& 2 (a_1^2 + a_2^2 + a_3^2) + 2 (a_1^2 a_2^2 + a_1^2 a_3^2 + a_2^2 a_3^2) - (a_1^4 + a_2^4 + a_3^4) - \sqrt{\delta} - 1, \\
    \delta ={}& \prod_{\mu, \nu=0}^{1} \left[(a_1 + (-1)^{\mu} a_2 + (-1)^{\nu} a_3)^2 - 1 \right] .
\end{align}
\end{subequations}
\end{widetext}

Finally, we can also compute the tripartite entanglement in Eq.~\eqref{eq:monogamy}. 
Since the global state is pure, the EoF between a partition \(A_i{:} A_jA_k\) is simply the reduced R\'enyi-2 entropy of \(A_i\). 
That is, 
\begin{equation} \label{eq:bi-entanglement-stdform}
    \mathcal{E}( \mathsf{A}_i{:}\mathsf{A}_j \mathsf{A}_k ) = \frac{1}{2} \ln a_i^2 .
\end{equation}
Whence, Eq.~\eqref{eq:monogamy} reduces to
\begin{equation}
    \mathcal{E}_2( \mathsf{A}_i; \mathsf{A}_j{:}\mathsf{A_k} ) = \frac{1}{2} \ln( \frac{a_i^2}{ g_j g_k } ) ,
\end{equation}
\bibliographystyle{apsrev4-2}
\bibliography{PhD-DickeProject.bib, lib_gabriel}

\end{document}